# Evaluation of new radio occultation observations among small satellites at Venus by data assimilation

Yukiko Fujisawa[1], Norihiko Sugimoto[1,2], Chi Ao[3], Asako Hosono[4], Hiroki Ando[5], Masahiro Takagi[5], Itziar Garate Lopez[6] and Sebastien Lebonnois[7]

1) Research and Education Center for Natural Sciences, Keio University, Yokohama 223-8521, Japan
2) Department of Physics, Keio University, Yokohama 223-8521, Japan
3) Jet Propulsion Laboratory, California Institute of Technology, Pasadena, CA 91109, USA
4) Department of Precision Engineering, The University of Tokyo, Tokyo 113-8656, Japan
5) Faculty of Science, Kyoto Sangyo University, Kyoto 603-8555, Japan
6) Bilboko Ingeniaritza Eskola (UPV/EHU), 48013 Bilbao, Spain
7) Laboratoire de Meteorologie Dynamique (LMD/IPSL), Sorbonne Universite, ENS, PSL Research University, Institut Polytechnique de Paris, CNRS, Paris, France

**Abstract:** We conducted observing system simulation experiments (OSSEs) for radio occultation measurements (RO) among small satellites, which are expected to be useful for future Venus missions. The effectiveness of the observations based on realistic orbit calculations was evaluated by reproduction of the "cold collar", a unique thermal structure in the polar atmosphere of Venus. Pseudo-temperature observations for the OSSEs were provided from the Venus atmospheric GCM in which the cold collar was reproduced by the thermal forcing. The vertical temperature distributions between 40 and 90 km altitudes at observation points were assimilated. The result showed that the cold collar was most clearly reproduced in the case where the temperature field in high-latitudes was observed twice a day, suggesting that the proposed observation is quite effective to improve the polar atmospheric structure at least. Although the cold collar was also reproduced in the OSSEs for Longwave Infrared Camera (LIR) observations, the result seemed unrealistic and inefficient compared to that obtained in the RO OSSEs. The present study shows that the OSSEs can be used to evaluate observation plans and instruments in terms of reproducibility of specific atmospheric phenomena, and applied to future missions targeting planetary atmospheres.

Y. Fujisawa, Research and Education Center for Natural Sciences, Keio University, 4-1-1, Hiyoshi, Kouhoku, Yokohama, Kanagawa, 223-8521, Japan. (yukiko@gfd-dennou.org)

**1. Introduction**

The atmosphere of Venus has a global, thick cloud cover at about 48–70 km altitude. Therefore, it is very difficult to optically observe atmospheric phenomena in and below the cloud. One of the most useful methods to observe the atmospheric structure in the cloud layer is the radio occultation (RO). On Venus, it enables us to obtain vertical temperature profiles in altitudes of approximately 40–90 km (e.g., *Tellmann et al.* 2009, *Ando et al.*, 2020). The RO observations have been performed in some Venus missions, such as the Pioneer Venus, Venus Express, and Akatsuki missions. Disadvantages of the RO measurements are that the frequency and coverage of the observations were strictly limited due to the communication environment between the orbiters and the Earth. In the case of Akatsuki, the frequency is less than once a week. A method to overcome such restrictions is the inter-satellite RO using multiple small satellites, which enables us to obtain vertical temperature profiles in a wide altitude range including the cloud layer globally and at high frequency. On Earth, such kind of observations have already been conducted by a fleet of small satellites designed to track radio signals transmitted by GNSS (Global Navigation Satellite System) satellites [*Kursinski et al.*, 1997; *Ho et al.*, 2020], but it has never been done on any planet other than Earth, including Venus.

Observation system simulation experiment (OSSE) is a method for testing a virtual observation system using data assimilation, which enables us to evaluate a planned observation in terms of reproducibility of a target phenomenon. The data assimilation estimates the most probable state using observation data and a forecast model, and in the OSSE, pseudo-observation data (nature run) generated by a well-known model, which is usually different from the forecast model, is assimilated as the observation data. On Earth, the OSSE is actively used for objective evaluation of meteorological forecasting skill when new observations including satellite observations are introduced [*Masutani et al.*, 2010]. However, there are few applications of the OSSE for planetary satellite observations. So far, we have performed the OSSEs targeting the reproduction of the cold collar for RO observations [S*ugimoto et al.*, 2019b], the thermal tide for Longwave Infrared Camera (LIR) observations [*Sugimoto et al.*, 2022b], and the equatorial Kelvin wave for Ultraviolet Imager (UVI) observations [*Sugimoto et al.*, 2021a, 2022a], and as far as we know, these are the only OSSEs applied to the Venus atmosphere. Even for Mars, the planet with the most active numerical modeling and observational research, the OSSE framework has been just proposed in recent years [*Reale et al.*, 2021]. Since a planetary satellite mission is limited in time and resources, it would be quite useful to investigate its effectiveness in advance using the OSSE to optimize its instruments and observation plans.

To implement the OSSE, we need both a data assimilation system and a reliable meteorological model which can reproduce the atmospheric phenomena realistically. We have developed a Venus GCM named AFES-Venus [*Sugimoto et al.*, 2014a], which is based on AFES (Atmospheric GCM For the Earth Simulator) optimized for the Earth simulator, and succeeded in reproducing the realistic super rotation [*Sugimoto et al.*, 2014b], the polar vortex [*Ando et al.*, 2016], the thermal tide [*Takagi et al.*, 2018], the planetary-scale streak structure [*Kashimura et al.*, 2019], and the gravity waves generated from the thermal tide [*Sugimoto et al.*, 2021b]. Using AFES-Venus and Local Ensemble Transform Kalman Filter (LETKF), we have also developed the first data assimilation system for the Venus atmosphere named ALEDAS-V (AFES LETKF Data Assimilation System for Venus) [*Sugimoto et al.*, 2017, 2019a]. In addition to the OSSEs mentioned above, we have constructed objective analysis by assimilating horizontal wind data obtained from UVI observations by Venus Express and Akatsuki. [*Sugimoto et al.*, 2019a, *Fujisawa et al.*, 2022].

*Sugimoto et al.* [2019b] performed the OSSE using ALEDAS-V, and assimilated idealized observations simulating RO observations among multiple small satellites, and examined how the reproduced atmospheric structure depends on frequency and coverage of the observations. They showed that if an ideal observation system was constructed for the polar region, it would be possible to reproduce the "cold collar", which is a cold latitudinal band surrounding the warm polar vortex at about 65 km altitude and 60°–80° latitudes [*Taylor et al.*, 1980]. The cold collar is a unique thermal structure closely related to the fast zonal wind (super rotation) and the mean meridional circulation at the cloud top levels [*Ando et al.*, 2016]. Therefore, the reproducibility of temperature and wind distributions in the polar region with a focus on the cold collar can be regarded as a barometer of realistic simulations of the Venus atmospheric dynamics.

In *Sugimoto et al.* [2019b], however, there was a large temperature bias between the model and the observation, and the observation points were unrealistically fixed for simplicity. In this study, as an extended work of *Sugimoto et al.* [2019b], we assimilate observations using more realistic conditions by simulating the RO measurements among small satellites with realistic orbits for a longer period, and introducing bias correction. In addition, the OSSE for LIR observations is also performed for the purpose of a comparison experiment to evaluate the usefulness of RO observations. We show the reproducibility of the temperature and wind distributions associated with the cold collar from the viewpoint of atmospheric dynamics.

## 2. Setup of experiments

### 2.1 Data assimilation system

The LETKF, which is used in the data assimilation system ALEDAS-V, is a kind of data assimilation method based on the Ensemble Kalman Filter (EnKF) [*Eversen*, 1994], and is a practical method with particularly excellent parallel computing efficiency. It is widely used in the Earth meteorology [*Miyoshi et al.*, 2007; *Yamazaki et al.*, 2017] and has also been applied to the Martian atmosphere [*Greybush et al.*, 2012]. The model variance of forecast model is given by the variability provided by the ensemble computation of forecast model. The estimates that minimize the error variance of the forecast model and the observation are obtained by the technique of the Ensemble Transform Kalman Filter [*Bishop et al.*, 2001], which is an application of the Kalman Filter [*Kalman*, 1960]. Furthermore, localization is adopted for computational efficiency, i.e., only observations within the specified horizontal and vertical distances are considered [*Ott et al.*, 2004].

In the present study, the number of ensemble members is 31. The inflation is fixed to 10 % which is to prevent the ensemble error from becoming too small due to model bias. The localization parameters are 400 km in horizontal and $\log P = 0.4$ in vertical, where *P* is pressure. The time interval of the data assimilation is set to 6 hours. The four-dimensional LETKF comprised 7-hour time slots at each analysis, and the observations are assimilated every hour depending on their availability.

ALEDAS-V uses AFES-Venus for ensemble forecasts. AFES-Venus is a full nonlinear dynamical

GCM based on the assumption of hydrostatic balance, designed for the Venus atmosphere [*Sugimoto et al.*, 2014a]. In this study, the horizontal resolution is set to T42, where T denotes the triangular truncation number for spherical harmonics; there are 128×64 horizontal grids at each level. The model atmosphere extends from the flat ground to 120 km, which is divided into 60 layers with constant thickness of 2 km. The model includes vertical eddy diffusion with a constant coefficient of 0.15 m$^2$ s$^{-1}$. Horizontal eddy viscosity is represented by the second-order hyper viscosity with a damping time for the maximum wave number component of 0.1 Earth days. At the lowest level, Rayleigh friction with a damping time of 0.5 Earth days is employed to take the surface friction into account. In the upper atmosphere above 80 km, a sponge layer with damping times gradually decreasing with height is applied only to the eddy components. Convective adjustment is used to suppress convective instability. The solar heating is based on Pioneer Venus observations [*Tomasko et al.*, 1980]. The infrared radiative process is simplified by a Newtonian cooling scheme, whose coefficients are based on *Crisp* [1986]. The temperature is relaxed to a prescribed horizontally uniform temperature distribution based on the Venus International Reference Atmosphere (VIRA) [*Seiff et al.*, 1985]. Details are described in our previous research [*Sugimoto et al.,* 2014a, b].

We use an idealized super rotating flow in solid-body rotation as an initial state of the velocity field, in which the zonal wind increases linearly with height from the ground to 70 km and reaches 100 m s$^{-1}$ at the equator. The initial temperature distribution is in gradient wind balance with the zonal wind. The direction of the planetary rotation is assumed to be eastwards (positive). Note that this is the same direction as the rotation of Earth, which is the opposite to the actual rotation of Venus. We have confirmed that a fully developed super rotating flow could be obtained from a motionless state with a small vertical viscosity excluding the thermal tide [*Sugimoto et al.,* 2019c], and the zonal mean zonal winds reproduced from the different two initial conditions with the super rotating flow and a motionless state converged to a similar state. However, since an extremely long time (~ thousands of Earth years) is needed to reproduce the fast super rotating flow from the motionless state, we use the idealized super rotating flow as the initial state in the present study. Using this initial state, we perform nonlinear numerical simulations for more than four Earth years as a spin-up. Then datasets sampled at 8-hour intervals in the quasi-equilibrium state are used as initial conditions for each 31-member ensemble for the ensemble run in the LETKF data assimilation.

**2.2 Idealized pseudo-observation data**

Idealized pseudo-observation data are provided from the Institut Pierre Simon Laplace Venus atmospheric GCM (IPSL-VGCM) [*Garate-Lopez and Lebonnois*, 2018] (nature run; NR), which has 96×96×50 total grid points, though the resolution differs from that of AFES-Venus. The assimilation period is 90 Earth days (0.77 Venus days). This is three times longer than the experimental period used in *Sugimoto et al.* [2019b]. Fig. 1a and 1b show latitude-height cross sections of zonal mean temperature in 40–90 km altitudes. In AFES-Venus without assimilation (free run forecast; FR), a structure similar to the cold collar, in which the poles are warmer than the surrounding latitudes, can be seen in 70–80 km on both hemispheres, but it is not seen near 67 km where satellite observations show the cold collar (Fig. 1a). On the other hand, in NR, the cold collar is clearly reproduced around 67 km in latitudes poleward of about 60°N(S) (Fig.1b). Fig. 1c show vertical profiles of the horizontal mean temperature in 40–90 km altitudes. There is a large difference in the horizontal mean temperatures between FR (blue line) and the original output of IPSL-VGCM (NR-org) (green line). The temperature difference is particularly large in 65–80 km altitudes, which is 13.2 K at 67 km altitude. This large temperature bias can break the circulation fields associated with the cold collar and super-rotations during long-term assimilation experiments. For this reason, we adjust horizontal mean temperature of the NR-org to be the same as FR at each altitude, and adopt it as nature run (NR) in this study. Note that NR-org was used for nature run in S*ugimoto et al.* [2019b].

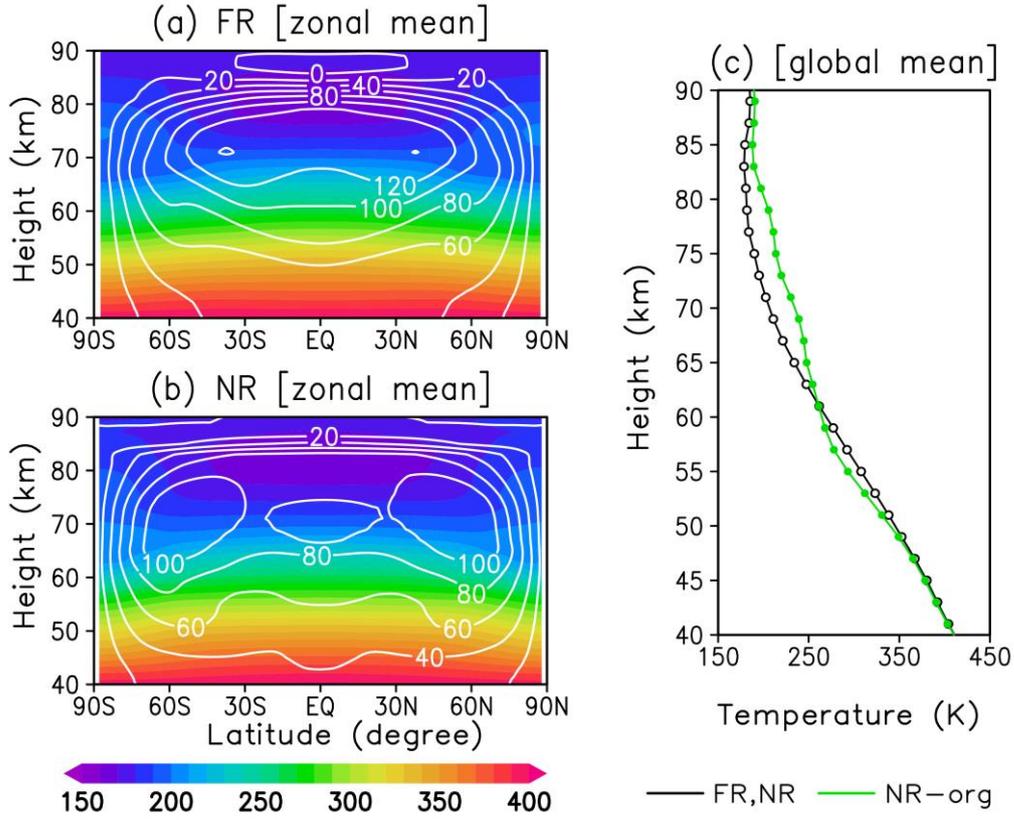

Fig. 1. Latitude-height cross sections zonal-mean temperature (color, K) and zonal-mean zonal wind (contour, m s$^{-1}$) in 40–90 km altitudes obtained for the cases of (a) FR and (b) NR, and vertical profiles of the horizontal mean temperature obtained for FR and NR (black line), and NR-org (green line). Results are averaged for the last 30 Earth days in the experimental period (90 Earth days; 0.77 Venus days). The NR (c) is corrected each altitude so that it is equal to the horizontal mean temperature of FR. The units are K.

## 2.3 Observation conditions

We assimilate vertical temperature distributions between 40 and 90 km altitudes at observation points (longitudes and latitudes), which are determined by assuming RO observations among three satellites (see below for details). The observation error is set to 3 K for simplicity. The temperature measurement error of the RO observation is basically determined by the stability of the oscillator and temperature retrieval error (initialization of temperature in the hydrostatic integration and assumption of spherical symmetry). It is expected from the oscillators installed in Akatsuki and Venus Express that the temperature measurement errors are ~ 0.1 K [*Imamura et al.*, 2017] and total error would be ~1–2 K including all error sources. On the other hand, in the present study, in order to prevent the forecasts from getting too close to the observations and thus ensure some variability among the ensemble members, random errors of 3 K are added to each observation. In the previous works [*Sugimoto et al.*, 2017], we have conducted experiments with different observation errors (6 K) and confirmed that there was almost no effect on the results.

The observation points of the RO observations in this study are realistically determined by simulating the orbits of three small satellites and checking for occultation opportunities between distinct pairs of satellites. We assume that all the satellites are in polar circular orbits (90° inclinations). The orbit altitudes of satellites A, B, and C are 350, 1500, and 3000 km, and the true anomalies (the angle between the direction of periapsis and the current position of the satellite, as seen from the main focus of the ellipse orbit of satellite) are 0°, 45°, and 90° at initial (Fig. 2). Fig. 3a and 3b show the observation points. P1, P2, and P3 represent the observation points obtained from the RO by satellites A and C, A and B, and B and C. Sum of P2 and P3 is represented by P2P3, and so on. In this paper, we use P1 and P3 as the observation cases with two satellites, and P1P2P3 as the case with three satellites. See SI for other cases. Table 1 shows the observation frequency at each observation point. P1, P3 and P1P2P3 have 12.4, 5.7, and 24.8 observations per day. The numbers of observation points in the polar regions (latitudes poleward of 75°N), which are considered to be important for the reproduction of cold collar, are 1.0, 0.4, and 2.1 times per day for P1, P3, and P1P2P3. In this study, the assimilations are performed for these 3 cases of observation points.

The horizontal temperature distribution near the cloud top have been obtained widely and frequently by LIR onboard Akatsuki [*Taguchi et al.*, 2012; *Kouyama et al.*, 2019] and Visible and Infrared Thermal Imaging Spectrometer (VIRTIS) onboard Venus Express [*Piccioni et al.*, 2007]. LIR can observe in low latitudes and VIRTIS can observe in middle and high latitudes continuously. These observations can be made at any latitude depending on the orbit of the satellite. In this study, we also perform the OSSE assuming LIR observations covering the polar region and compare the result with those obtained for the RO OSSEs. Since LIR observes thermal radiation emitted from 45 to 85 km altitudes [*Taguchi et al.*, 2007], the LIR observations strongly depend on the cloud structure and there is uncertainty in observation altitude. Although the height of clouds varies greatly with latitude [*Haus et al.*, 2014], in this experiment, for simplicity, we assume that the LIR senses the temperature field at a fixed altitude of 65 km for assimilation. The observation area is 8 AM–4 PM local time in the daytime 0°–90°N, and the LIR observations are assimilated twice a day. This observation area is based on the previous OSSE performed for Akatsuki LIR in *Sugimoto et al.* [2022b]. While *Sugimoto et al.* [2022b] covers the equatorial region, the observation area is shifted to the northern hemisphere in this experiment for a comparison. The horizontal resolution of LIR observations is the same as NR (96×96).

Table 2 shows the total number of observation data per day for the LIR and P1P2P3 (RO) experimental cases. In the P1P2P3 case, 28 observation data are obtained in the vertical direction per an observation point. Since RO observations are performed 24.8 times a day, the total number of observation data per day is $24.8 \times 28 = 694.4$. In the LIR case, the observation data in the daytime 8 AM–4 PM local time (31 points) and 0°–90°N (49 points) can be obtained in the horizontal direction. Since LIR observations are performed 2 times a day, the total number of observation data per day is $49 \times 31 \times 2 = 3038.0$. If they are simply compared, the total number of LIR observation data is more than four times larger than that of the P1P2P3 data. However, since the zonal spacing of the observation points is very narrow in the polar regions, it remains unclear how the observation data are effective.

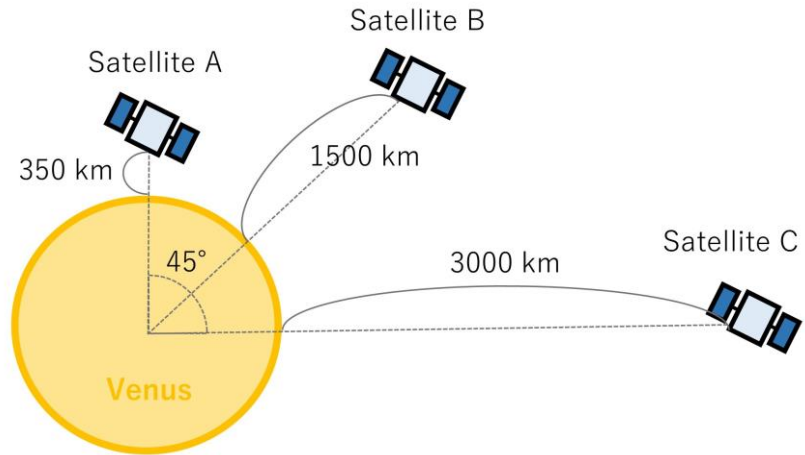

Fig. 2. Positional relationship of satellites in RO measurements between satellites using multiple small satellites. The initial state in the orbits simulations is shown.

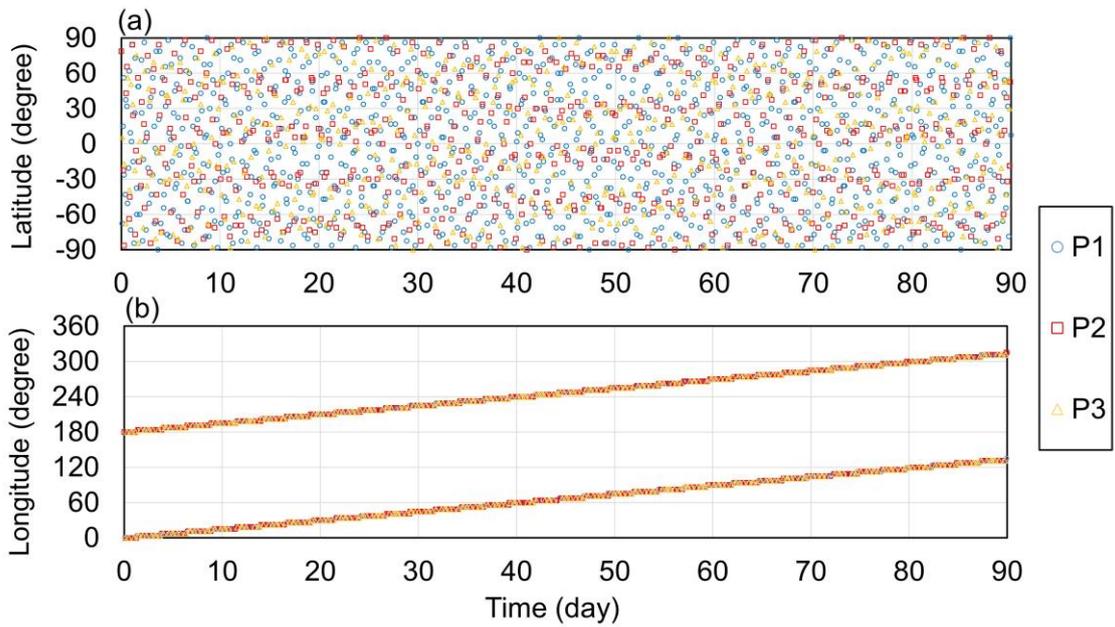

Fig. 3. (a) Latitude-time cross sections and (b) Longitude-time cross sections of observation points. P1 (Blue Circles): Observation points obtained from satellites A and C, P2 (Red Squares): Observation points obtained from satellites A and B, and P3 (Yellow Triangles): Observation points obtained from satellites B and C.

| Experimental case / Area | P1 | P3 | P1P2P3 |
|---|---:|---:|---:|
| **Global** | 12.4 | 5.7 | 24.8 |
| **North of 75 deg.** | 1.0 | 0.4 | 2.1 |

Table 1. Experimental cases and observation frequency. The units of frequency are times per day. Headings in the vertical column indicate regions of observation and headings in the horizontal row indicate experimental cases (P1, P3 and P1P2P3).

| Experimental case / Area | P1P2P3 | LIR |
|---|---:|---:|
| **Global** | 694.4 | 3038.0 |
| **North of 75 deg.** | 58.8 | 558.0 |

Table 2. Experimental cases and total number of observation points. The units of total number are times per day. Headings in the vertical column indicate regions of observation and headings in the horizontal row indicate experimental cases (P1P2P3 and LIR).

## 3. Results

### 3.1 OSSEs of RO

Fig. 4 shows temperature distributions at 67 km altitude on the northern hemisphere obtained for 10, 50, and 90 Earth days from the start of experiments. Movies of the temporal variation of 30 Earth days from the start of experiments are also shown in SI. During the experiment in NR, difference in the zonal mean temperatures between 87.9°N and 71.2°N latitudes is maintained around 10 K; the polar region is warmer than its surroundings, showing that the cold collar is reproduced (Fig. 4a, f, k). On the other hand, in FR, the temperature monotonically decreases with latitude, i.e., the cold collar is not reproduced (Fig. 4b, g, l). In the RO OSSEs, on the 10 Earth days from the start of experiments, the temperature difference is 8.4 K in P1P2P3 with the highest observation frequency (Fig. 4e), 3.9 K in P1 with the second highest (Fig. 4c), and –1.8 K in P3 with the lowest (Fig. 4d). These results suggest that there is a clear correlation between the observation frequency and the temperature difference. On the 50 Earth days, the temperature differences are 9.7 K and 10.1 K in P1P2P3 and P1, respectively, while it remains negative at –0.9 K in P3. On the 90 Earth days, the temperature difference is positive in all the three cases. Note that, however, no correlation was found between the observation frequency and the temperature difference at the 90 Earth days. The temperature difference is 3.1 K in P1P2P3, 2.0 K in P1, 3.8 K in P3 (Fig. 4m, n, o). Interestingly, it is the largest in P3 with the lowest observation frequency. In P1P2P3, although the temperature is high near the north pole, it is also high in the surrounding region (71.2°N) (Fig. 4o). As a result, the temperature difference is reduced in this case.

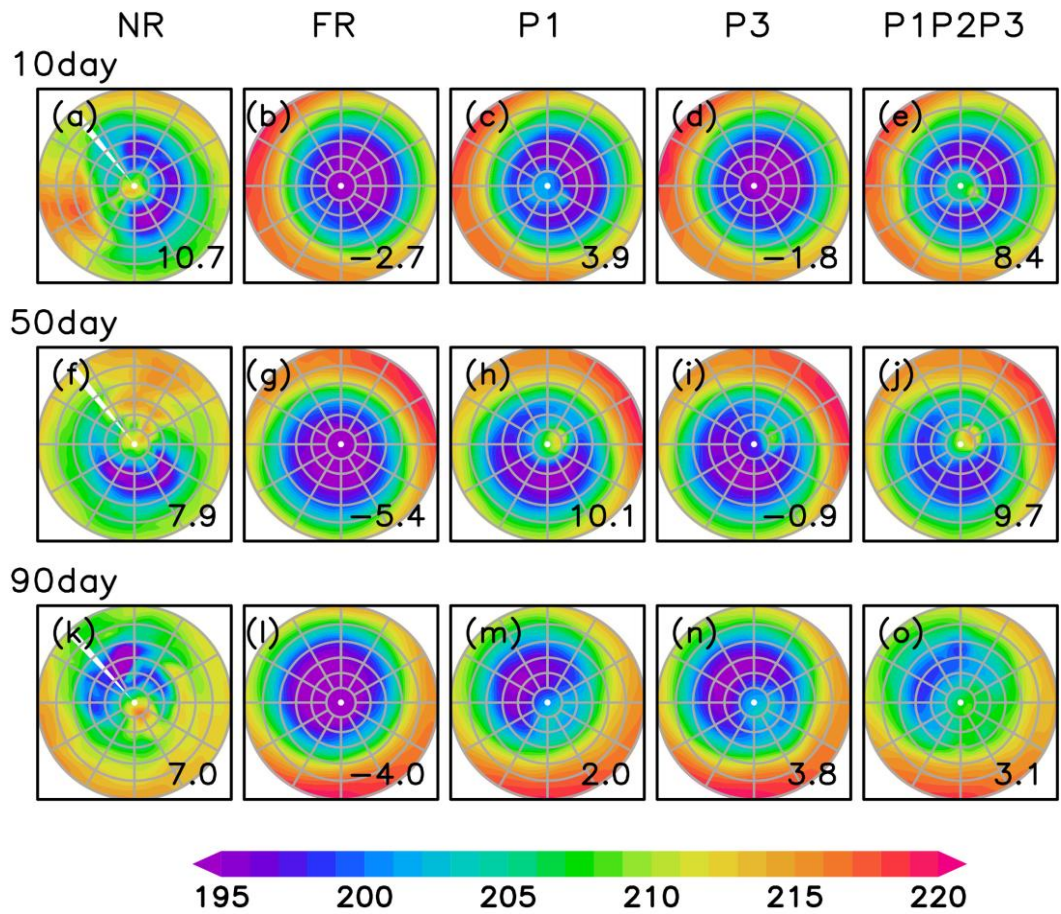

Fig. 4. Temperature at 67 km altitude in 30°–90°N latitudes seen from the north pole for (a, f, k) NR (nature run), (b, g, l) FR (free run), (c, h, m) P1, (d, i, n) P3, (e, j, o) P1P2P3. Results are shown for (a, b, c, d, e) 10, (f, g, h, i. j) 50 and (k, l, m, n, o) 90 Earth days from start of the experiment. The units are K. The number on the right-bottom corner in each panel is the zonal mean temperature difference between 87.9°N and 71.2°N latitudes.

Fig. 5a shows the time series of the zonal mean temperature difference between 87.9° and 71.2°N latitudes. As shown in Fig.4a, f, and k and Fig. 4b, g, and l, in NR (black line), the temperature difference is always positive, i.e., the pole is warmer than its surroundings, and the cold collar is reproduced, in contrast to that the temperature difference is always negative in FR (red line). In P1P2P3 (light blue line) with the highest observation frequency, the temperature difference turns positive on 5 Earth days after the start of experiment, and reaches almost the same value as obtained in NR after 10 Earth days. Although the temperature difference becomes smaller than that of NR from around 55–75 Earth days, it remains roughly equivalent to NR. In P1 (green line) with the next highest observation frequency, the temperature difference increases more slowly than in P1P2P3, but it catches up with that in NR after 25 Earth days. It becomes lower than that in NR around 40–50 Earth days, but it is roughly the same as that in NR for other periods. In P3 (blue line) with the lowest observation frequency, the temperature difference turns positive around 23 Earth days and catches up with that in NR on 75 Earth days. These results suggest that the higher the observation frequency is, the faster the cold collar could be reproduced. However, after the cold collar is reproduced, it seems that the temperature differences obtained in the RO OSSEs are close to that in NR, and do not depend on the observation frequency.

In order to quantify the effect of observations, root-mean-square-deviation (RMSD) expressed by the following equation (1) is calculated at 67 km altitude:

$$RMSD = \sqrt{\frac{1}{N}\sum_{i=1}^{N}(X_i - x_i)^2}, \tag{1}$$

where $X_i$ are physical values of NR, $x_i$ are those with the assimilation (analysis) for each case, and $N$ is a total number of horizontal grid points. The RMSD shows the difference between NR and the assimilation result, and smaller RMSD indicates that the data assimilation result is closer to NR.

Fig. 5b shows the time series of RMSD of temperature at 67 km altitude in 0°–90°N latitudes. In FR (red line), the RMSD fluctuates with an amplitude of 5–6 K during the experiment. In P1P2P3 (light blue line) with the highest observation frequency, it decreases to 3 K at 90 Earth days. This result means that the temperature field in P1P2P3 gets closer to that in NR than that in FR. In P1 (green line) and P3 (blue line), the RMSDs are 3.5 K and 4 K, respectively. Fig. 5b indicate that the RMSD could correlate positively with the observation frequency.

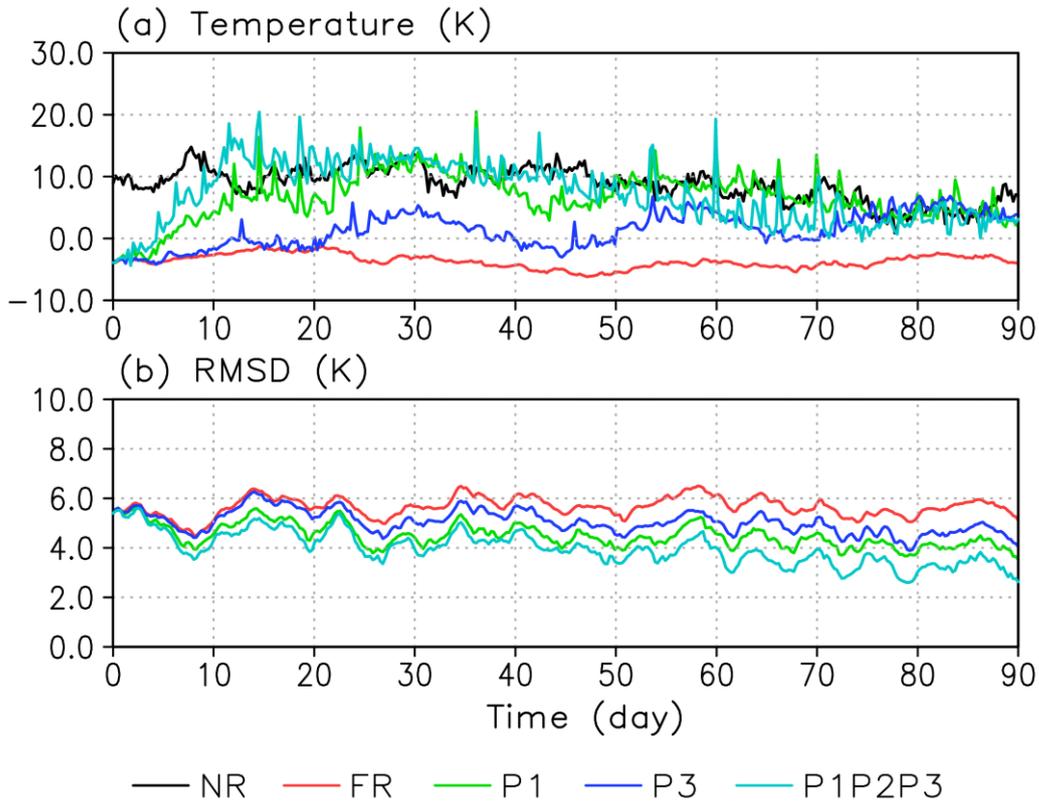

Fig. 5. (a) The zonal-mean temperature difference from 87.9 to 71.2°N and (b) root-mean-square-deviation (RMSD) of temperature at 67 km altitude in 0°–90°N latitudes for NR (black), FR (red), P1 (green), P3 (blue) and P1P2P3 (light blue). The units are K.

Fig. 6 shows latitude-height cross sections of the deviations of zonal-mean zonal wind and zonal-mean temperature from those in NR. The temperature difference between FR and NR is negative in latitudes poleward of 50°N and altitudes of 65–70 km, and its magnitude increases with latitude toward the pole (Fig. 6a). On the other hand, in P1, P3, and P1P2P3, this negative temperature difference is small, i.e., the temperature fields become close to that in NR (Fig. 6b, c, d). This means that the temperature fields are improved in altitudes where the cold collar exists. In P1P2P3 with high observation frequency, the temperature difference almost vanishes in this region (Fig. 6d). It is confirmed that the temperature field is more improved in these altitudes by the more frequent observations. In addition, in P1P2P3, the temperature difference is also small in the region where it is positive in latitudes of 0°–50°N. This is consistent with the small temperature difference between the pole and its surroundings, as shown in Fig. 4o.

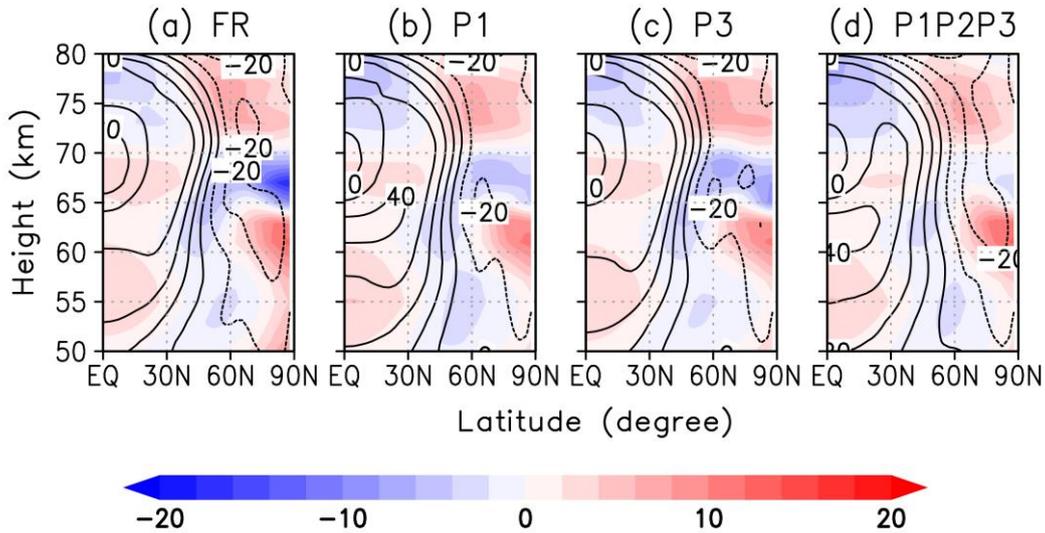

Fig. 6. Latitude-height cross sections of the deviations from NR of zonal-mean zonal wind (contour) and zonal-mean temperature (color) for (a) FR, (b) P1, (c) P3 and (d) P1P2P3. Results are shown for the average of 60–90 Earth days from the start of the experiment. The units are m s$^{-1}$ for wind and K for temperature, respectively. The contour intervals are 10 m s$^{-1}$.

In order to examine the changes in temperature and circulation fields due to the assimilation in more detail, the differences in temperature and zonal wind from FR are calculated (Fig. 7). In P1, there is a large temperature difference around the pole at 67 km altitude, and the region with a temperature difference greater than 4 K extends equatorward to around 50°N (Fig. 7a). The zonal wind is significantly accelerated by more than 10 m s$^{-1}$ in the high latitudes below this level. Similar changes in the zonal wind field are also seen in P3 and P1P2P3 (Fig. 7b, c). In these experiments for the RO observations, even though only the temperature is assimilated, the wind fields are also changed accordingly in the mid-latitudes and polar region.

The zonal-mean zonal wind (so-called super-rotation) in low latitudes at 70–80 km altitude decelerates due to the assimilation. In P1P2P3, which has the highest observation frequency, the zonal-mean zonal wind decelerates by about 12 m s$^{-1}$ (Fig. 7c). These changes are consistent with that the super-rotation in NR (Fig. 1d) is about 10–20 m s$^{-1}$ slower than that in FR [*Garate-Lopez and Lebonnois*, 2018] (Fig. 1b, d). The deceleration of zonal-mean zonal wind due to the assimilation of temperature observations could be explained by the thermal wind balance between the temperature and zonal wind fields, i.e., the zonal wind field could be changed by the temperature filed which is modified by the assimilation.

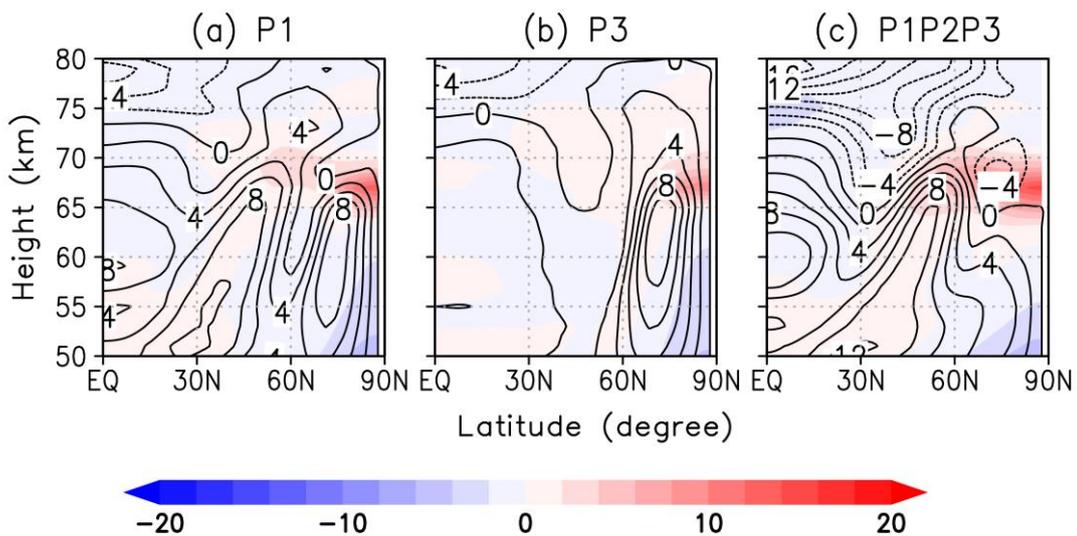

Fig. 7. Latitude-height cross sections of the difference from FR of zonal mean zonal wind (contour) and temperature (color) for (a) P1, (b) P3 and (c) P1P2P3. Results are shown for the average of 60–90 Earth days from the start of the experiment. The units are m s$^{-1}$ for wind and K for temperature, respectively. The contour intervals are 2 m s$^{-1}$.

*Ando et al.* [2016] and *Ando et al.* [2023] investigate how the cold collar depends on the polar circulation. *Ando et al.* [2016] conducted GCM experiments forced by the diurnal solar heating and the zonal-mean solar heating, and analysed the polar circulation. In the diurnal solar heating cases, the downward flow of the residual mean meridional circulation (RMMC) was enhanced in the polar region, and the warm polar region is formed by adiabatic heating due to the downward branch of RMMC.

To examine the RMMC in this experiment, meridional and vertical components of the RMMC, $\bar{v}^*$ and $\bar{w}^*$, are defined as follows [Andrews *et al.*, 1987]:

$$\bar{v}^* \equiv \bar{v} - \frac{1}{\rho_0}\frac{\partial}{\partial z^*}\left(\frac{\rho_0 \overline{v'\theta'}}{\partial \theta/\partial z^*}\right), \tag{2}$$

$$\bar{w}^* \equiv \bar{w} + \frac{1}{R_V \cos\varphi}\frac{\partial}{\partial \varphi}\left(\frac{\overline{v'\theta'}\cos\varphi}{\partial \theta/\partial z^*}\right), \tag{3}$$

where overbars denote the zonal average, primes are the deviation from the zonal average, v and w are meridional and vertical velocity, respectively, $\theta$ is potential temperature, $R_V$ is the Venus radius, $\rho_0$ is the atmospheric density, $\varphi$ is latitude, and $z^* = -H \log(p/p_s)$ is log-pressure height with $H$ being the scale height.

Fig. 8 shows the latitude-altitude distribution of the zonal-mean temperature and RMMC. In all the RO OSSEs (P1, P3, and P1P2P3), the upward (downward) flow appears in latitudes of 0°–30°N (40°–90°N), and the poleward flow is more enhanced in higher altitudes.

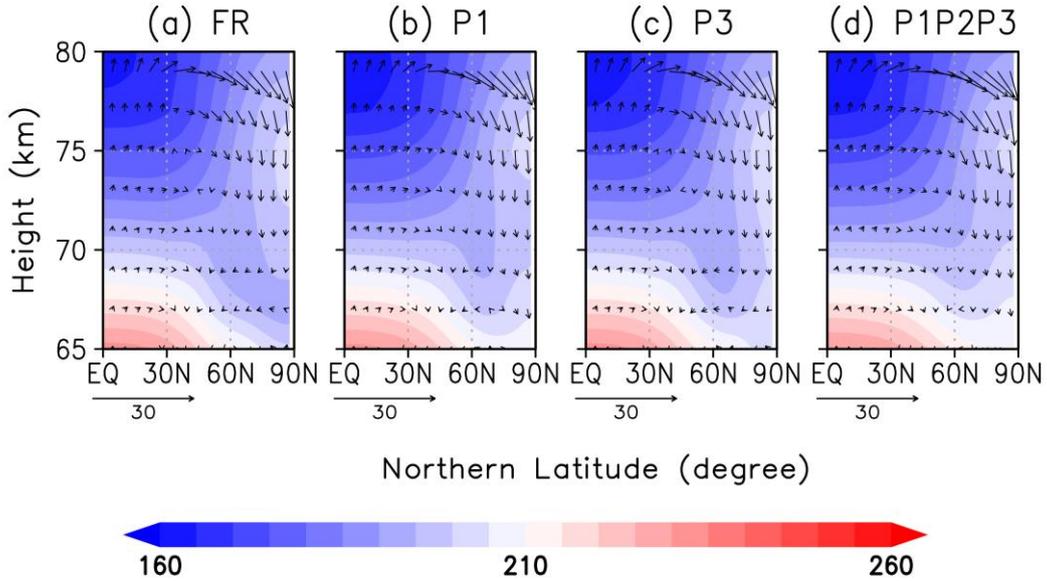

Fig. 8. Latitude-height distributions of zonal-mean RMMC (vector; m s$^{-1}$) and temperature (color; K) at 66–80 km altitudes obtained in (a) FR, (b) P1, (c) P3, and (d) P1P2P3. Results are shown for the average of 60–90 Earth days from the start of the experiment. $\bar{w}^*$ is multiplied by $10^3$.

To clarify the difference of RMMC between the assimilation experiments and FR, Fig. 9 show the latitude-altitude distributions of the zonal-mean meridional $\bar{v}^*$ and vertical $\bar{w}^*$ components of RMMC. In P1P2P3, the meridional flow is stronger than that in FR around 60°N above 75 km altitude, with a maximum of 4 m s$^{-1}$ (Fig. 9d). The vertical downward flow is also stronger in latitudes poleward of 60°N by about 30 mm s$^{-1}$ than that in FR (Fig. 9h), which is enhanced over the wide range of 65–80 km altitudes. These results indicate that the RMMC in P1P2P3 is more enhanced and extends to lower altitudes than that in FR. Compared to the RMMCs obtained in P1 and P3, the most enhanced is that of P1P2P3 with the highest observation frequency.

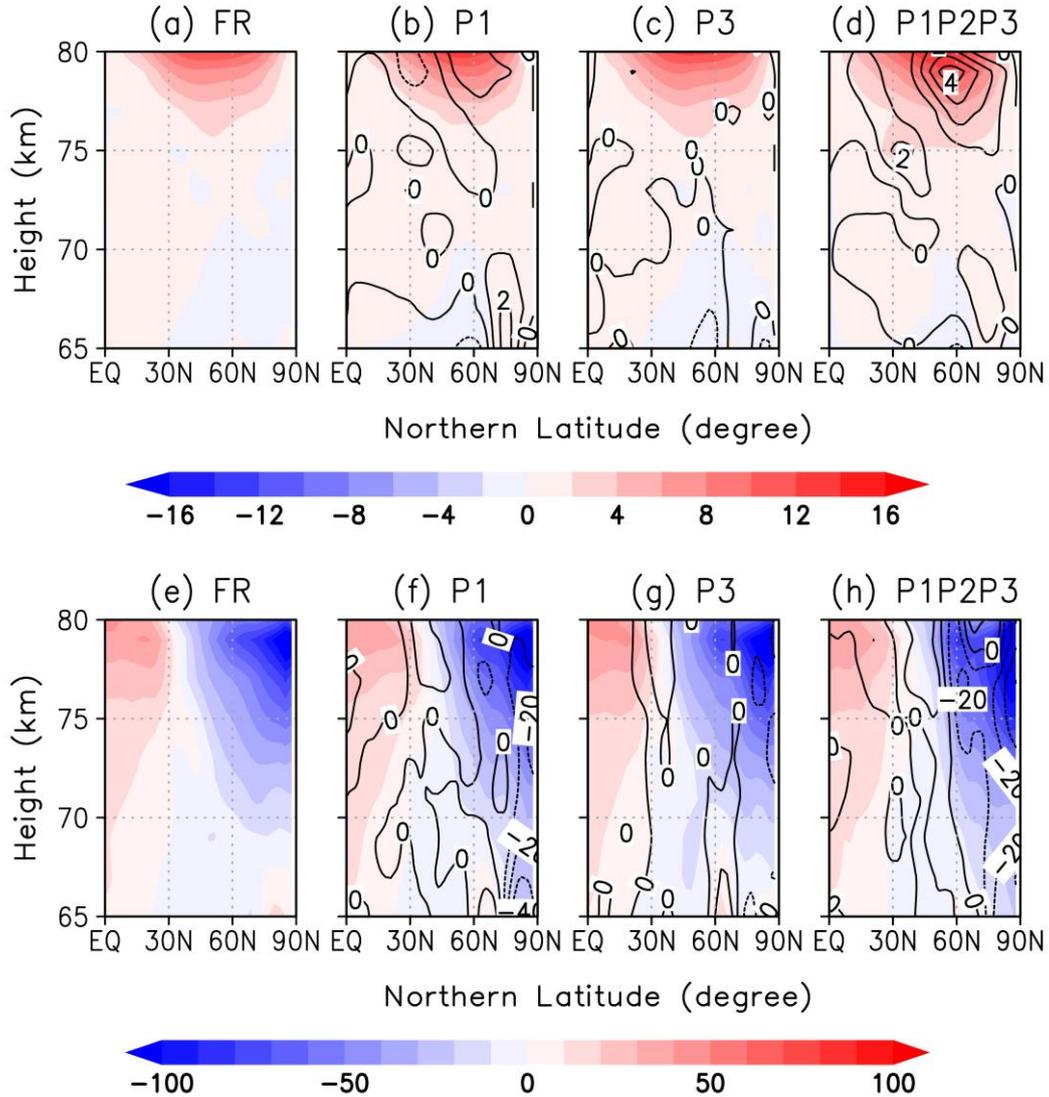

Fig. 9. Latitude-height distributions of the zonal-mean meridional and vertical component of RMMC ($\bar{v}^*$, $\bar{w}^*$) at 65–80 km altitudes obtained in (a, e) FR, (b, f) P1, (c, g) P3, and (d, h) P1P2P3. Solid (dashed) black contours represent positive (negative) differences with FR. (a, b, c, d) Color is $\bar{v}^*$, and red (blue) color represents a poleward (equatorward) flow. The interval of black contour is 1 m s$^{-1}$. (e, f, g, h) Color is $\bar{w}^*$, and Red (blue) color represents an upward (downward) flow.

The interval of black contour is 10 mm s$^{-1}$. It should be noted that $\bar{w}^*$ is multiplied by $10^3$.

## 3.2 OSSE of LIR

Fig. 10 shows temperature distributions in latitudes of 30°–90°N at 67 km altitude on the northern hemisphere obtained in the LIR OSSE. The difference of the zonal-mean temperatures between 87.9°N and 71.2°N latitudes are 9.5, 8.4, and 0.4 K at 10, 50, and 90 Earth days (Fig. 10a, b, c). As shown in Fig. 10d, the temperature difference in LIR (yellow line) becomes almost the same as in NR (black line) after about 50 Earth days if we ignore fast fluctuations in it (Fig. 10d). However, the temperature difference fluctuates unnaturally in a short period of time, which is – 13.5 K at 80 Earth days and 26.9 K at 88 Earth days. These fluctuations are quite different from those in NR (black line) (see also SI for the movie). Fig. 10e shows that the RMSD in LIR (yellow line) behaves like that in P1P2P3 (light blue line). Its magnitude is 3 K in P1P2P3 (light blue line), and 3.8 K in LIR (yellow line) at 90 Earth days, indicating that the temperature field in P1P2P3 is closer to that in NR.

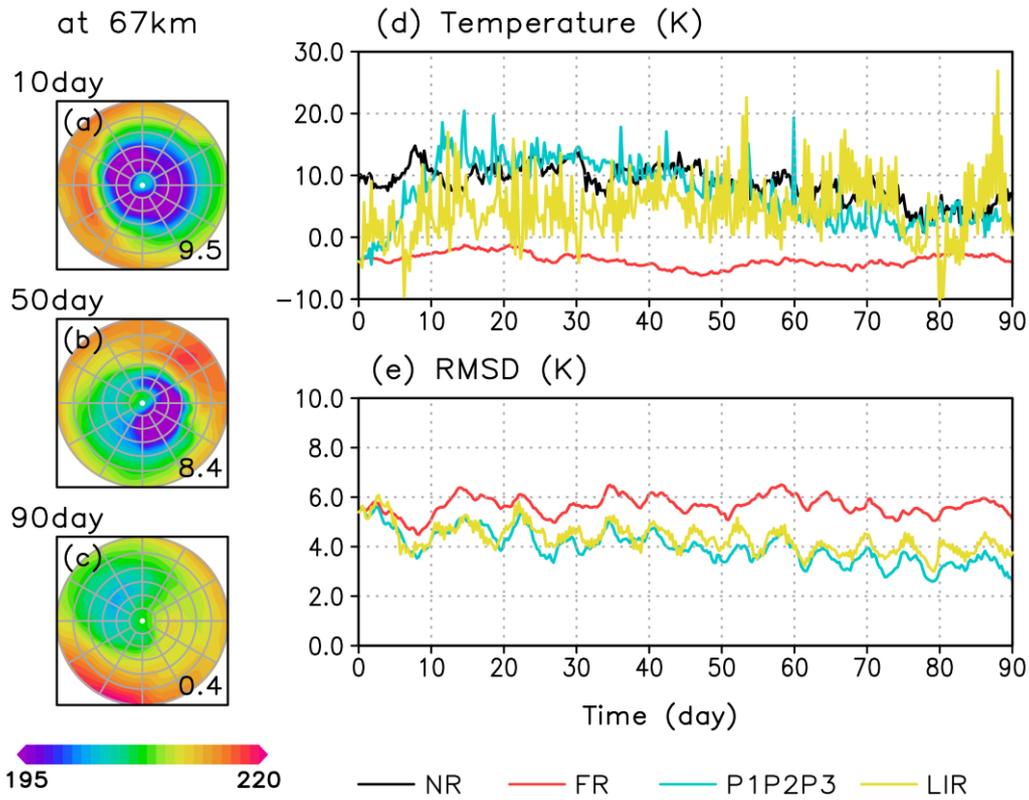

Fig. 10. The LIR OSSE case. (a, b, c) Temperature at 67 km altitude in 30°–90 °N latitudes seen from the north pole. Results are shown for (a) 10, (b) 50 and (c) 90 Earth days from start of the experiment. The units are K. The number on the right-bottom corner in each panel is the zonal mean temperature difference between 87.9 °N and 71.2 °N latitudes. (d) The zonal-mean temperature difference from 87.9 to 71.2°N and (c) root-mean-square-deviation (RMSD) of temperature at 67 km altitude in 0°–90°N latitudes for NR (black), FR (red), P1P2P3 (light blue) and LIR (yellow). The units are K.

Fig. 11 shows latitude-height cross sections of the deviations of zonal-mean zonal wind and zonal-mean temperature from those in NR and FR. In the northern hemisphere where the assimilation is performed, the temperature difference between LIR and NR is negative at 65–70 km altitude (Fig.

11a), but its magnitude is smaller than FR (Fig. 6a). This means that the temperature field is improved in altitudes where the cold collar exists. In the difference between LIR and FR, the zonal wind is decelerated and accelerated alternately in 60–70 km altitudes; it is decelerated by 20 m s$^{-1}$ around 40°N and 80°N, and accelerated by 4 m s$^{-1}$ around 60°N (Fig. 11b). The magnitude of the zonal wind change in LIR is larger than that in the RO assimilations of P1, P3, and P1P2P3 (Fig. 7a, b, c). The zonal wind in low latitudes and 70–80 km altitudes is decelerated by more than 20 m s$^{-1}$, which is larger than that in P1P2P3 (Fig. 7c).

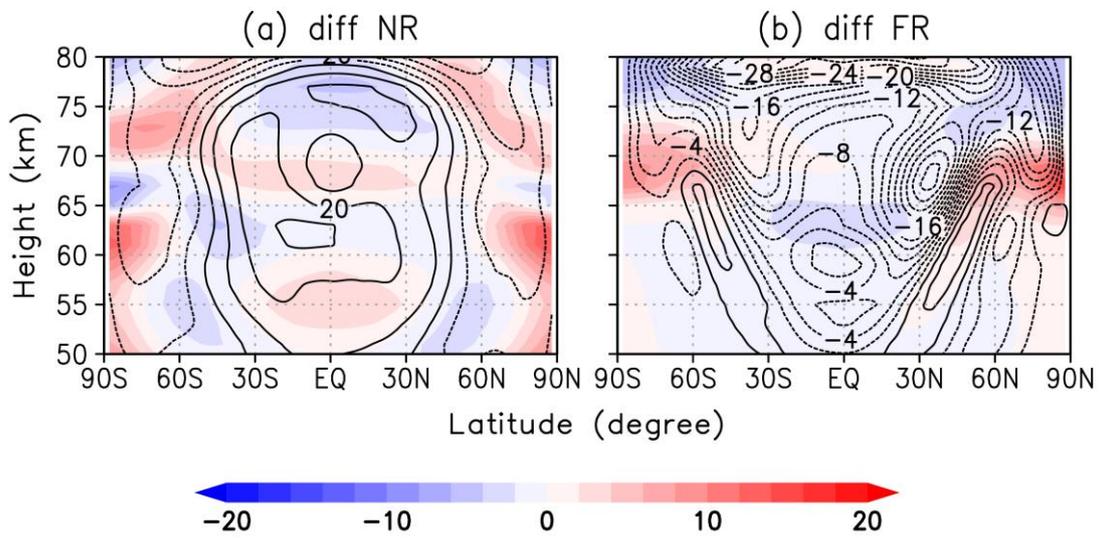

Fig. 11. The LIR OSSE case. (a) Latitude-height cross sections of the deviations from NR of zonal-mean zonal wind (contour) and zonal-mean temperature (color). Results are shown for the average of 60–90 Earth days from the start of the experiment. The units are m s$^{-1}$ for wind and K for temperature, respectively. The contour intervals are 10 m s$^{-1}$. (b) Latitude-height cross sections of the difference from FR of zonal mean zonal wind (contour) and temperature (color). Results are shown for the average of 60–90 Earth days from the start of the experiment. The units are m s$^{-1}$ for wind and K for temperature, respectively. The contour intervals are 2 m s$^{-1}$.

Fig. 12a shows the latitude-altitude distribution of zonal-mean temperature and RMMC obtained in the LIR OSSE. As in the other experiments, the upward (downward) flow appears in latitudes of 0°–30°N (40°–90°N). Fig. 12b and c show the latitude-altitude distributions of the zonal-mean meridional $\bar{v}^*$ and vertical $\bar{w}^*$ components of RMMC. The meridional flow is stronger than that in FR by about 2 m s$^{-1}$ in mid-latitudes above 75 km altitude (Fig. 12b), which is smaller than the meridional flow of 4 m s$^{-1}$ in P1P2P3 (Fig. 9d). On the other hand, the vertical flow is stronger than that in FR intensified with some points below –100 mm s$^{-1}$ (Fig. 12c) at polar region, which is stronger than –30 mm s$^{-1}$ in P1P2P3 (Fig. 9h). In addition, the vertical flow is weak around 50° and 80°N, and is strong at 60°N at an altitude of 80 km. These striped structures appear to be paired with the zonal wind (Fig. 11b).

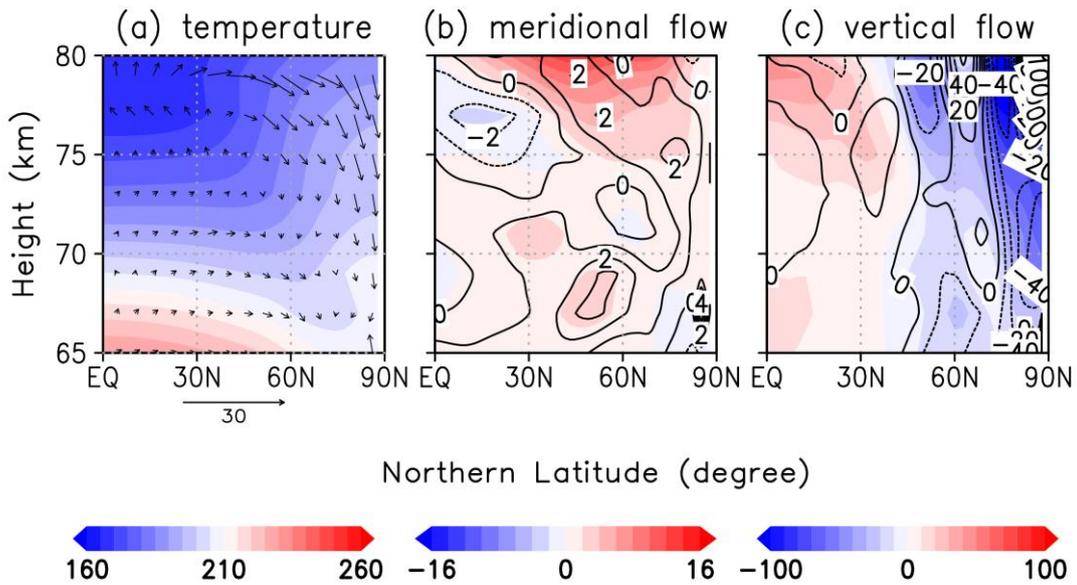

Fig. 12. The LIR OSSE case. (a) Latitude-height distributions of zonal-mean RMMC (vector; m s$^{-1}$) and temperature (color; K). (b, c) Latitude-height distributions of the zonal-mean meridional and vertical component of RMMC ($\bar{v}^*$, $\bar{w}^*$) at 65–80 km altitudes. Solid (dashed) black contours represent positive (negative) differences with FR. (b) Color is $\bar{v}^*$, and red (blue) color represents a poleward (equatorward) flow. The interval of black contour is 1 m s$^{-1}$. (c) Color is $\bar{w}^*$, and Red

(blue) color represents an upward (downward) flow. The interval of black contour is 10 mm s$^{-1}$. It should be noted that $\bar{w}^*$ is multiplied by 10$^3$. Results are shown for the average of 60–90 Earth days from the start of the experiment.

## 4. Summary and discussion

In this study, we conducted OSSEs assuming RO observations among multiple small satellites and LIR observations with a focus on the reproducibility of the cold collar. For the RO OSSEs, the observation points are obtained based on realistic orbit calculations. In the case of three satellites, the vertical profile of temperature in the polar regions is observed at about two points per day, that is, about once every 12 hours. In this case, the cold collar could be reproduced by consecutive observations over about 5 Earth days or more. The temperature difference between the pole and its surroundings showed that the correlation between the observation frequency and the temperature difference decreases as the experiment period increases. On the other hand, the RMSD showed that the temperature field becomes more similar to that obtained in NR as the observation frequency increases. In the zonal mean field, it is found that the zonal wind (super-rotation) is also corrected, although only the temperature field is assimilated. The zonal wind is accelerated around the pole at altitudes where the cold collar reproduced, and moderately decelerated in the upper cloud layer at low latitudes. Since the zonal wind derived from cloud images of Akatsuki observations [*Horinouchi et al.*, 2018] is slower than that obtained in FR, it could be determined that the zonal wind field has been improved as well as the temperature field by assimilating the temperature observation. The RMMC suggests that the assimilation enhances the polar downward flow associated with the cold collar; that is, the polar region is warmed by the adiabatic heating due to the downward flow, as proposed by *Ando et al.* [2016]. These results suggest that the RO observations are useful for improving the atmospheric thermal structures such as the cold collar, and possibly the global atmospheric circulation at wide altitudes of 40–90 km in the numerical model.

In *Sugimoto et al.* [2019b], the cold collar was reproduced most clearly when the vertical profile of temperature was observed at 2 to 3 points once every 4 to 6 hours in the polar region, and it can also be reproduced for lower observation frequency, which was 3 points once every 24 hours. The experiment period in their study is 30 Earth days, which is shorter than that in the present experiments. First of all, the present experiments show that the RO observations by multiple small satellites orbiting Venus, which can be achieved with currently available technology, instead of fixed-point observations such as *Sugimoto et al.* [2019b], can reproduce the cold collar. The RO observations by multiple small satellites have recently been considered as one of the new Venus missions [*Yamamoto et al.*, 2021]. Secondly, the cold collar was still successfully reproduced, although the observation frequency was considerably lower than that assumed in *Sugimoto et al.* [2019b], which was about once every 12 hours in the most frequently observed case (P1P2P3) and once every 60 hours in the least frequently observed case (P3). Finally, long-term experiments showed that the lower the observation frequency, the longer it takes to reproduce cold collar, and the correlation between the observation frequency and the cold collar reproduction decreases as the experiment period increases. Note that for long-term experiments, it is important to maintain the basic atmospheric field of GCM by bias correction. For the reproduction of the cold collar by assimilation, it might be effective to assimilate the temperature gradient in the latitudinal direction rather than assimilating the temperature itself obtained from the observation.

In the LIR OSSE, which was conducted for a comparison with the RO OSSEs, a northern dayside region at the top of the cloud layer is observed once every 12 hours. In this case, as with the RO OSSEs, the pole regions were warmer than their surroundings. However, the temperature difference between the pole and its surroundings unnaturally fluctuates between positive and negative values in a short period of time. The zonal-mean zonal wind and RMMC show that the circulation field associated with the cold collar is enhanced compared to the RO OSSEs. In addition, the vertical component of RMMC is weak around 50° and 80°N, and is strong at 60°N at an altitude of 80 km. Such striped structures are also weakly shown in the RO OSSEs. It is our future work to investigate whether they are real structures on Venus or assimilation-induced shocks.

It is difficult to judge which is better, the RO or LIR observations, to reproduce the cold collar based on these OSSEs only. The RO data could be used more easily for the assimilation, because RMSD and the correlation between observation frequency and the temperature difference between the poles and their surroundings are somewhat straightforward. In addition, the assimilation of RO observation is more efficient, since the LIR observation requires four times as many data as the RO observation to obtain an equivalent result. Launching small satellites for RO has generally less cost than a large satellite for the LIR instrument. These economic conditions cannot be ignored for actual planetary missions.

A data assimilation system such as ALEDAS-V can be used to perform several types of OSSEs for the Venus atmosphere [e.g., *Sugimoto et al.*, 2019b, 2021a, 2022a, 2022b]. The OSSEs can be also used to optimize satellite orbits and observation plans by evaluating reproducibility of various atmospheric structures such as the cold collar, the planetary-scale streak, and the planetary-scale waves, and enable us to make more effective future observation plans in the planetary atmospheric sciences. It should be noted that dynamics of various phenomena would be elucidated by the data assimilation regardless of results from OSSEs or analysis. Furthermore, we can elucidate dynamics of various phenomena from the results of the data assimilation (e.g., RMMC), regardless of OSSEs or analysis.

## Acknowledgments

This study was conducted under the joint research project of the Earth Simulator Center with title "High resolution general circulation simulation of Venus and Mars atmosphere using AFES". The work is partly supported by MEXT | Japan Society for the Promotion of Science (JSPS) grants JP19H01971, JP19H05605, JP20K04064. This study was supported by GSC (Global Science Campus) of JST (Japan Science and Technology Agency). Work performed by C. Ao was carried out at the Jet Propulsion Laboratory, California Institute of Technology, under a contract with the National Aeronautics and Space Administration. Results from the simulation experiments performed in this paper are available at Mendeley Data (http://dx.doi.org/10.17632/cg8nxtkf94.4).